\documentclass[number,preprint,review,12pt]{elsarticle}  
\usepackage{amsmath} 
\usepackage{booktabs} 
\usepackage{tabularx} 
\usepackage{hyperref} 
\usepackage{subcaption}
\newtheorem{lemma}{Lemma}
\newtheorem{theorem}{Theorem}
\newtheorem{assumption}{Assumption}
\newproof{proof}{Proof}
\def\bel{\begin{lemma}}
\def\eel{\end{lemma}}
\date{}

\usepackage{amssymb}

\journal{}

\begin{document}

\begin{frontmatter}



\title{Two-sample test of sparse stochastic block models}


\author{Qianyong Wu and Jiang Hu$^*$ }
\cortext[mycorrespondingauthor]{Corresponding author}

\address{School of Mathematics $\&$ Statistics, Northeast Normal University, Changchun, China}

\begin{abstract}
The paper discusses a statistical problem related to testing for differences between two sparse networks with community structures. The community-wise edge probability matrices have entries of order $O(n^{-1}/\log n)$, where $n$ represents the size of the network. The authors propose a test statistic that combines a method proposed by Wu et al. \cite{WuTwoSampleSBM2022} and a resampling process. They derive the asymptotic null distribution of the test statistic and provide a guarantee of asymptotic power against the alternative hypothesis. To evaluate the performance of the proposed test statistic, the authors conduct simulations and provide real data examples. The results indicate that the proposed test statistic performs well in practice.

\end{abstract}




\end{frontmatter}


\section{Introduction}\label{sec1}

Networks have been widely used to describe relationships between individuals or interactions between units of complex systems in diverse fields, such as biology, computer science, sociology and  many others \cite{jiCoauthorshipCitationNetworks2016,gangrade2019efficient,ghoshdastidarTwosampleHypothesisTesting2020,pontesBiclusteringExpressionData2015,westveldMixedEffectsModel2011,bickelAsymptoticNormalityMaximum2013}.  With the development in data collection and the explosion of social media network analysis has become an increasingly important part of data analysis.

Given a large network, one of the important task is to find its community structure. For example, identifying communities of customer's network can help promoting a brand of product \cite{krishnareddyGraphBasedApproach2002}. In the protein-protein network, communities are groups of proteins that carry specific functions in the cell, finding communities can help us understanding the organization and dynamics of cell functions \cite{chen2006detecting}. In the Coauthorship and Citation networks for statisticians, communities detection can help us finding the envelopment of research habits, trends and topological patterns of statisticians \cite{jiCoauthorshipCitationNetworks2016,jiCocitationCoauthorshipNetworks2021}.
The stochastic block model \cite{hollandStochasticBlockmodelsFirst1983} is a very popular model for community structures in network data and has been extensively studied in computer science and statistics literature. The stochastic block model assigns $n$ nodes to one of $K$ disjointed blocks. Given the memberships, the observed network is  recorded by the symmetric adjacency matrix $A$. The entries $A_{ij}(i > j)$  of the symmetric adjacency $A$ are independent
Bernoulli random variables, where the parameter $E(A_{ij})$ depends only on the memberships of nodes $i$ and $j$.

Community detection is recovering the membership upon observing a single observation of $A$. This problem has received considerable attention from different research areas, many methods have been proposed such as spectral clustering\cite{jinFastCommunityDetection2015,maSemisupervisedSpectralAlgorithms2018,leiConsistencySpectralClustering2015}, likelihood methods\cite{aminiPseudolikelihoodMethodsCommunity2013,bickelAsymptoticNormalityMaximum2013,newmanMixtureModelsExploratory2007} and modularity maximization\cite{newmanFindingCommunityStructure2006}, see \cite{abbeCommunityDetectionStochastic2018} for a review. However, most of these methods assume we know the number of clusters which we do not. 
For example, the spectral clustering and its variants estimate the community membership by applying $k$-means clustering to the rows of the matrix formed by the $K_0$ leading singular vectors of $A$. Here $K_0$ is the hypothetical number of communities and should be given at first.
Most of these methods take the number of clusters $K$ as known a priori but in practice, $K$ is often unknown. To address this problem, a large number of methods have been proposed, including likelihood-based methods \cite{wangLikelihoodbasedModelSelection2017,saldanaHowManyCommunities2017a}, random matrix theory-based methods \cite{leiGoodnessoffitTestStochastic2016, bickelHypothesisTestingAutomated2016,le2022estimating,dongSpectralBasedHypothesis2020}, network cross-validation \cite{liNetworkCrossvalidationEdge2020,chenNetworkCrossValidationDetermining2018,huCorrectedBayesianInformation2020}, and the maximum entry-wise deviation approach \cite{huUsingMaximumEntryWise2020}, to name but a few. 

Analogous to single network problems, two-sample hypothesis tests for random networks arise naturally in neuroscience, social networks and
machine learning \cite{bassettHierarchicalOrganizationHuman2008,zhangDifferentialDependencyNetwork2009,chen2006detecting}. There has been several work on this important problem. Tang et al. proposed a  statistic for comparing two random dot product graphs by adjacency spectral embeddings in \cite{tangSemiparametricTwoSampleHypothesis2017,tangNonparametricTwosampleHypothesis2017}. Ghoshdastidar et al. in \cite{ghoshdastidarTwosampleHypothesisTesting2020} proposed two test statistics based on the estimates of the Frobenius norm and the spectral norm between the link probability matrices of the two samples,  and provided a clear characterization of the minimax separation threshold. Ghoshdastidar and von Luxburg \cite{ghoshdastidarPracticalMethodsGraph2018a}, Chen et al.\cite{chenHypothesisTestingPopulations2021} proposed test statistics based on the extreme eigenvalues of a scaled and centralized matrix,  and proved that the statistics 
approximately follow the Tracy-Widom law under the null hypothesis.

Most of the literature, however, focus on cases where the sample size is large, which limits the scope for inference. For example, omics data typically results in one large interaction network, that is, the sample size of either population is one. Hence, we need to design new two sample tests for a small sample case. 
Hu et al. \cite{huUsingMaximumEntryWise2020} proposed a  goodness-of-fit test based on the maximum entry of the adjacency matrix for the stochastic block model.
Inspired by the work of Hu et al. \cite{huUsingMaximumEntryWise2020}, Wu et al. \cite{WuTwoSampleSBM2022} proposed an approach for testing two dense random networks based on SBM, where the sample size is one. 

Early researches concern the dense networks, but their performance usually becomes unsatisfactory when the network becomes sparse. One of the reason is that sparse networks with small expected degrees fail to concentrate on their expectations. This is because such networks have vertices with degrees that are unusually high or low \cite{leConcentrationRegularizationRandom2017}. However, sparse networks are common in real-world scenarios such as social networks and gene co-expression networks, where the edges may not be easily observable despite containing hundreds of thousands of objects. It is noteworthy that sparse networks have proven to be useful in machine learning and have demonstrated their effectiveness in a variety of practical applications \cite{palCommunityDetectionSparse2021,roheSpectralClusteringHighdimensional2011,aminiPseudolikelihoodMethodsCommunity2013,jingCommunityDetectionSparse2021,WuTestingStochasticBlockModels2022}. In this project, we will propose a new method for conducting a two-sample test on sparse stochastic block models. 

The paper is organized as follows. In section \ref{sec2}, we explain why the method proposed by Wu et.al \cite{WuTwoSampleSBM2022} fails on sparse networks. We state the new test statistic with its asymptotic null distribution and asymptotic power in Section \ref{resample}.  
To illustrate the performance of the test statistic, some simulations and real data examples are given in Sections \ref{sec3} and \ref{rdata}, respectively.  Finally, we conclude this paper in Section \ref{conclusion}.
\section{Two sample test problem on dense network}\label{sec2}
Suppose that two  random networks over the same  vertices are given, and their adjacency matrices are generated from symmetric link probability matrix $P_1$ with $A_{1,ij}\sim$ $Bernoulli(P_{1,ij})$, $i, j = 1, 2, \ldots , n,$ and another sample of adjacency matrix $A_{2,ij}$ is generated with link probability matrix $P_2$.
Is there a difference between the two networks? The answer can be formulated in the following
hypothesis test framework:
\begin{align}\label{eq2.1} 
	H_0:P_1=P_2  \quad vs \quad H_1:P_1 \neq P_2 .
\end{align}
Assume that $A_1$ and $A_2$ have community structures, so they can be modeled by SBM. Consider a stochastic block model with $n$ nodes and $K$ communities, the probability mass function for the adjacency matrix $A_\ell$, $\ell=1,2,$ is
$$
P(A_\ell)=\prod\limits_{1 \leq i<j \leq n}B_{\ell,g_{i}g_{j}}^{A_{\ell, ij}}(1-B_{\ell,g_{i}g_{j}})^{1-A_{\ell,ij}}.
$$
where $g \in \left\{1,2,\ldots, K\right\}^{n}$ is the membership vector and $B_\ell \in [0,1]^{K\times K}$, $\ell =1, 2$ are the symmetric community-wise edge probability matrices.
As a result $A_1$ and $A_2$ can be  parameterized by $(g_1,B_1)$ and $(g_2,B_2)$, respectively.
In this regime, the hypothesis test (\ref{eq2.1}) becomes
\begin{align} \label{eq2.2} 
	H_0:B_1=B_2  \quad vs \quad H_1:B_1 \neq B_2 .
\end{align}
under the assumption that $g:=g_1 = g_2$.

Given the hypothetical number of communities $K$, let $\hat{g}$ be an estimated community membership vector. 
Define $\hat{n}_{k}= \#( \hat{V}_{k} )$, and $\hat{V}_{k}=\hat{g}^{-1}
(k)$, 
$1\leq k \leq K$. We consider the plug-in estimator of $B_\ell$, $\ell =1, 2$, which are proposed in \cite{leiGoodnessoffitTestStochastic2016}:
\begin{align} \label{eq2.3} 
	\hat{B}_{\ell,kl}=\left\{
	\begin{array}{rcl}
		\frac{\sum\nolimits_{i\in \hat{V}_{k},j\in \hat{V}_{l}}A_{\ell,ij}}{\hat{n}_{k}\hat{n}_{l}}, & & {k     \neq     l} ,   \\
		\frac{\sum\nolimits_{i,j\in \hat{V}_{k},i\leq j }A_{\ell,ij}}{\hat{n}_{k}(\hat{n}_{k}-1)/2},  &      & {k=l}.\\
	\end{array} \right.
\end{align}
To test the null hypotheses (\ref{eq2.1}) with a divergent $K$, Wu et al.\cite{WuTwoSampleSBM2022} proposed
the entry-wise deviation,
$$\hat\rho_{ik}=\frac{1}{\sqrt{ \#( \hat g^{-1}(k)\backslash\left\{ i\right\} )}}\sum_{j\in \hat g^{-1}(k)\backslash\left\{ i\right\}}\frac{A_{1,ij}-A_{2,ij}}{\sqrt{\hat B_{1, \hat g_{i}\hat g_{j}}(1-\hat B_{1,\hat g_{i}\hat g_{j}})+\hat B_{2,\hat \hat{g}_{i}\hat g_{j}}(1-\hat B_{2,\hat g_{i}\hat g_{j}})}},$$ where $\hat{g}^{-1}(k)=\{i:1\leq i\leq n, \hat{g}_i=k \}$,
and $\#(\hat{g}^{-1}(k)\backslash\left\{ i\right\})$ stands for the cardinality of set $\hat{g}^{-1}(k)\backslash\left\{ i\right\}$.

Denote 	$
L_n = \max\limits_{1 \leq i \leq n,1 \leq k \leq K} \vert \hat\rho_{ik} \vert,
$ under the dense network assumption in their paper, Wu et al. \cite{WuTwoSampleSBM2022} established the theoretical
properties of $L_n$. Their proposed test is not suitable for sparse networks, as discussed in their paper. Actually when the community-wise edge probability $\max\limits_{1 \leq i \leq K,1 \leq k \leq K}  B_{ij}=O(\frac{\log n}{n})$, $\hat\rho_{ik}$  can be expressed as a standard normal distribution $N(0, 1)$
plus an additional term $O_p(\frac{1}{\sqrt{\log n}})$. In this case the test statistic $L^{2}_{n}-2\log(2Kn)+\log\log(2Kn)$ converges to a Type-I extreme value distribution plus a term of
order $O_p(1)$. Thus, $L^{2}_{n}-2\log(2Kn)+\log\log(2Kn)$ can not converge to the Gumbel distribution under the sparse network setting,  which is true as Wu et al. \cite{WuTwoSampleSBM2022} showed under the dense network setting.

\section{Resampling method}\label{resample}
Several methods have been proposed to accommodate the sparsity of network. Jing et al. \cite{jingCommunityDetectionSparse2021}
proposed a method of partitioning sparse network using the symmetrized Laplacian inverse matrix (SLIM). Gao et al. \cite{gaoAchievingOptimalMisclassification2017} proved that normalized spectral clustering  can help improve the clustering result in sparse network. Wu et al. \cite{WuTestingStochasticBlockModels2022} developed a resampling method to test whether the sparse network can be adequately fitted by a stochastic block model with $K$ communities.

In this paper we will propose a novel test statistic by resampling $\hat{\rho}_{ik}$ and establish its theoretical
properties. Firstly, calculate $\hat{\rho}_{ik}$ ($1 \leq i \leq n,1 \leq k \leq K$)  based on Wu et al.  \cite{WuTwoSampleSBM2022} Secondly, for any fixed community $k \in \{1,2, \cdots ,K\} $ we randomly draw $S$ samples from $\{\hat{\rho}_{1k},\cdots \hat{\rho}_{nk}\}$ at the $m$-th realization $(m=1,\cdots,M)$, denote them as $\{\hat{\rho}_{m_1k},\cdots \hat{\rho}_{m_Sk}\}$.  Subsequently, calculate 
$$\hat{\gamma}_{mk}=\frac{\hat{\rho}_{m_1k}+\hat{\rho}_{m_2k}+\hat{\rho}_{m_Sk}}{\sqrt{S}}.$$
In this case, $\hat{\gamma}_{mk}$ can be expressed as a standard normal distribution $N(0, 1)$ plus an additional term $O_p(1/\sqrt{S\log n})$. One can see that the resampling process accelerates the speed of the additional term converge to 0.
Denote $F_n = \max\limits_{1 \leq i \leq n,1 \leq k \leq K} \vert \hat{\gamma}_{mk} \vert,$ intuitively large value of $F_n$ provides evidence to reject $H_0$  in (\ref{eq2.2}). For establishing the theoretical results, we need three assumptions.
\begin{assumption}\label{as:1}
There exists two positive constants $c_1 $ and $c_2$ such that $\frac{c_1n}{K}\leq \min\limits_{1 \leq k \leq K} n_k\leq \max\limits_{1 \leq k \leq K} n_k \leq \frac{c_2n^2}{K^2\log{^2} n}$ for all $n$.
\end{assumption}
\begin{assumption}\label{as:2}
For $\ell=1,2$ and $ 1\leq k,l \leq K$, $B_{\ell,kl} \in (0,1)$ and $\min\limits_{k,l} nB_{\ell,kl} \rightarrow \infty$, and $B_\ell$ have no identical rows, for $\ell=1,2$. 
\end{assumption}
\begin{assumption}\label{as:3}
Assume that $S = \max\{O(q^{-1/2}),o((n/\log n)^{c_3})\}$, where $q = \max\limits_{k,l}B_{\ell,kl}$ and
$K = o((n/\log n)^{c_3})$, for $0 < c_3 < 0.5$. $MK = o(n)$ and $\min\{n,M,S\}\rightarrow \infty$.	
\end{assumption}
Let $T :=T_n= L^{2}_{n}-2\log(MK)+\log\log(MK)$. 
Then we have the following asymptotic results for the test statistic $T$.
\begin{theorem}\label{the:2.1}
Suppose that Assumptions \ref{as:1}, \ref{as:2} and \ref{as:3} hold. Then under the null hypothesis $H_0:B_1=B_2$, as $n\to\infty$, we have that
\begin{align} \label{eq3.1} 
\lim\limits_{n\to\infty}P(T\leq y) = \exp\left\{-\frac{1}{\sqrt{\pi}}e^{-y/2}\right\}.
\end{align}
\end{theorem}
The proof of Theorem \ref{the:2.1} is collected in the supplementary materials.

It is worth noting that the right hand side of (\ref{eq3.1}) is the cumulative distribution function of the Gumbel 
distribution  with location parameter $\mu =-2\log(\sqrt{\pi})$ and scale parameter  $\beta=2$.
Therefore, 
	for the testing problem (\ref{eq2.2}),  we reject 
	$H_0$ when $T \geq t_{1-\alpha}$, where $t_{\alpha}$ is the $\alpha$th quantile of the Gumbel distribution 	$Gumbel(\mu ,\beta )$. 	
	
	Next, we consider the comparation for the tests based on $T$, which we summarize in the following theorem. Our work considers the sparse networks where the magnitude of the maximum expected node degree can be as small as $\log n$. 
	\begin{theorem}\label{the:2.2}
		Under the alternative hypothesis $H_1:B_1 \neq B_2$, suppose that \ref{as:1}, \ref{as:2} and \ref{as:3} hold and $\max\limits_{1 \leq i \leq n,1 \leq j \leq n} \vert B_{1,g_{i}g_{j}}-B_{2,g_{i}g_{j}} \vert  = \Omega(\sqrt{\frac{\log MK}{n}})$,  we have that as $n\to\infty$,
		$$ 
		P(T\ge c\log(MK))\rightarrow 1,
		$$
		for any positive constant $c$.	
	\end{theorem}
	The proof of Theorem \ref{the:2.2} is collected in the supplementary materials.
	
	\section{Numerical experiments}\label{sec3}
	In this section, we illustrate the performance of our proposed
	test statistic in simulated datasets. 
	Note that  in the following settings, we use the  symmetrized Laplacian inverse matrix (SLIM) algorithm in \cite{jingCommunityDetectionSparse2021} to estimate the label $\hat{g}$.

	\subsection{The null distribution}
	In this simulation, we examine the finite sample null distribution
	of the test statistic $T$ and verify the results in  Theorem \ref{the:2.1}. Assume that there are two 
	communities, the community-wise edge probability matrices $B_1=B_2=B$, with $B_{11} = B_{22 }= 7\frac{\log n}{n}$ and $B_{12 }= B_{21 }= 3\frac{\log n}{n}$.
	We consider a small network with $n = 200$. The membership vector $g$ is generated by sampling each entry independently from $\{1,2\}$ with equal probability.
	
\begin{figure}[h]%
		\centering
		\includegraphics[width=0.9\textwidth]{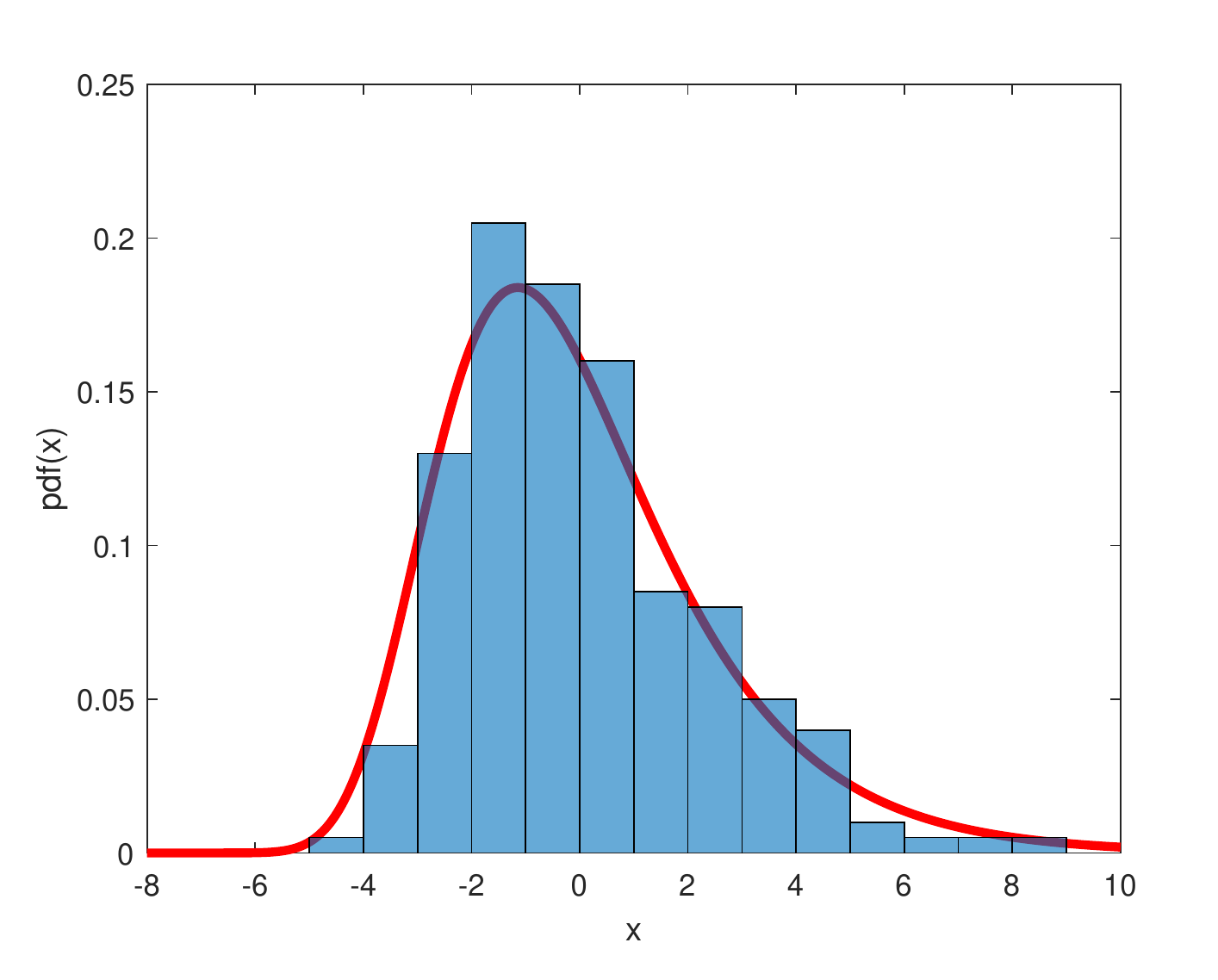}
		\caption{Histogram of the test statistic $T$ and the density function of the Gumbel distribution with $\mu =-2\log(\sqrt{\pi})$ and $\beta = 2$.}\label{fig1}
	\end{figure}

	In Figure \ref{fig1}  we present the histogram plot of our proposed test statistic from 1,000 independent replications under the null hypothesis. The theoretical limit distribution 
	(red line) is also plotted as a reference. It visually confirms the
	results in Theorem \ref{the:2.1}.
	
	\subsection{Type I and Type II errors under the dense networks}
	In this simulation, we compare the performance of our test statistic $T$ with the maximum entry-wise deviation statistic proposed by Wu et al. in \cite{WuTwoSampleSBM2022} and the largest eigenvalue statistic proposed by Ghoshdastidar et al. in \cite{ghoshdastidarPracticalMethodsGraph2018a} and Chen et al. in \cite{chenHypothesisTestingPopulations2021}, their test statistics are denoted by $T^{+}$ and $TW$ respectively.
	
\begin{figure}[h]
		\centering
		\begin{minipage}[t]{0.48\textwidth}
			\centering
			\includegraphics[height= 6cm,width=6cm]{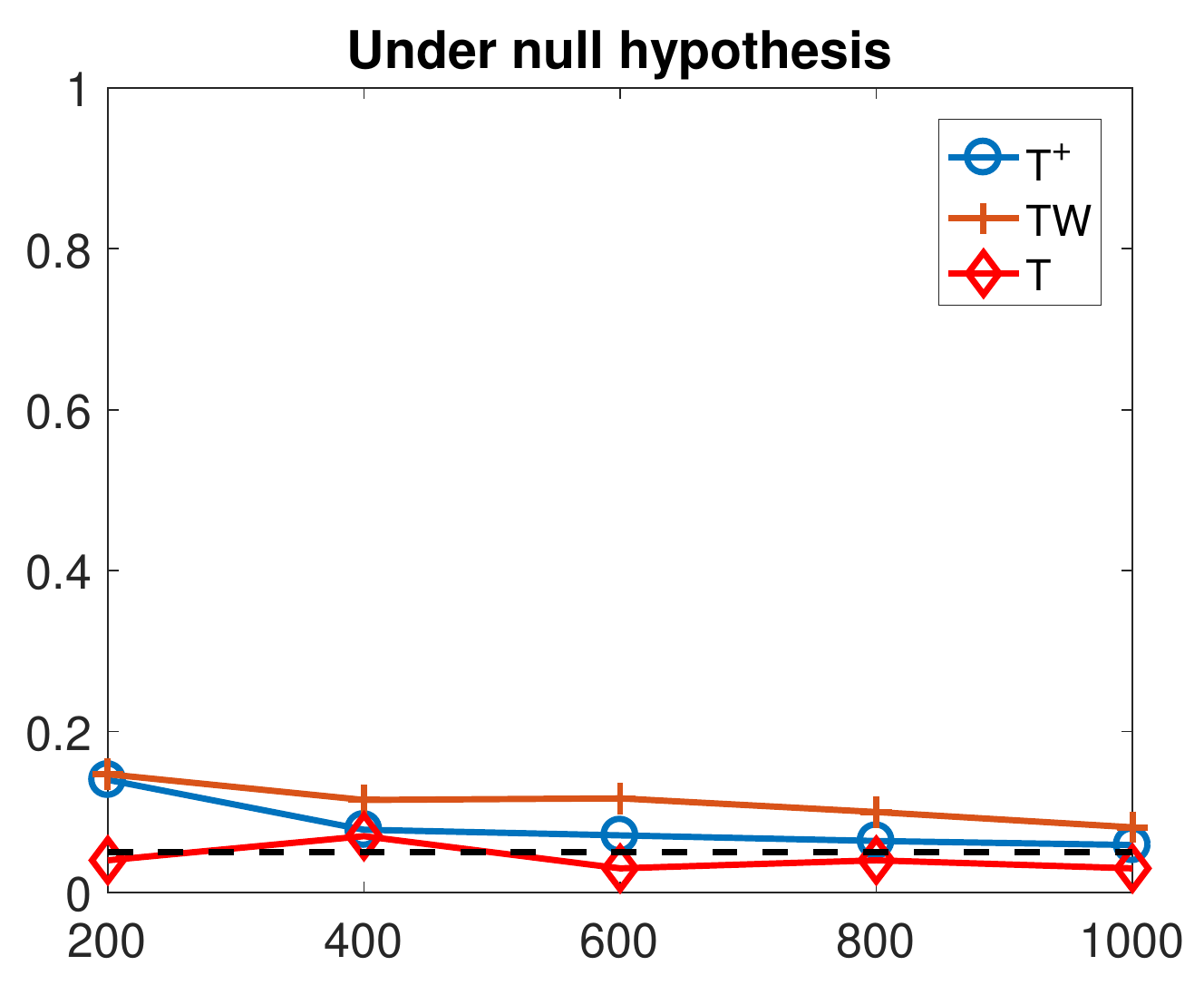}
			
		\end{minipage}
		\begin{minipage}[t]{0.48\textwidth}
			\centering
			\includegraphics[height= 6cm,width=6cm]{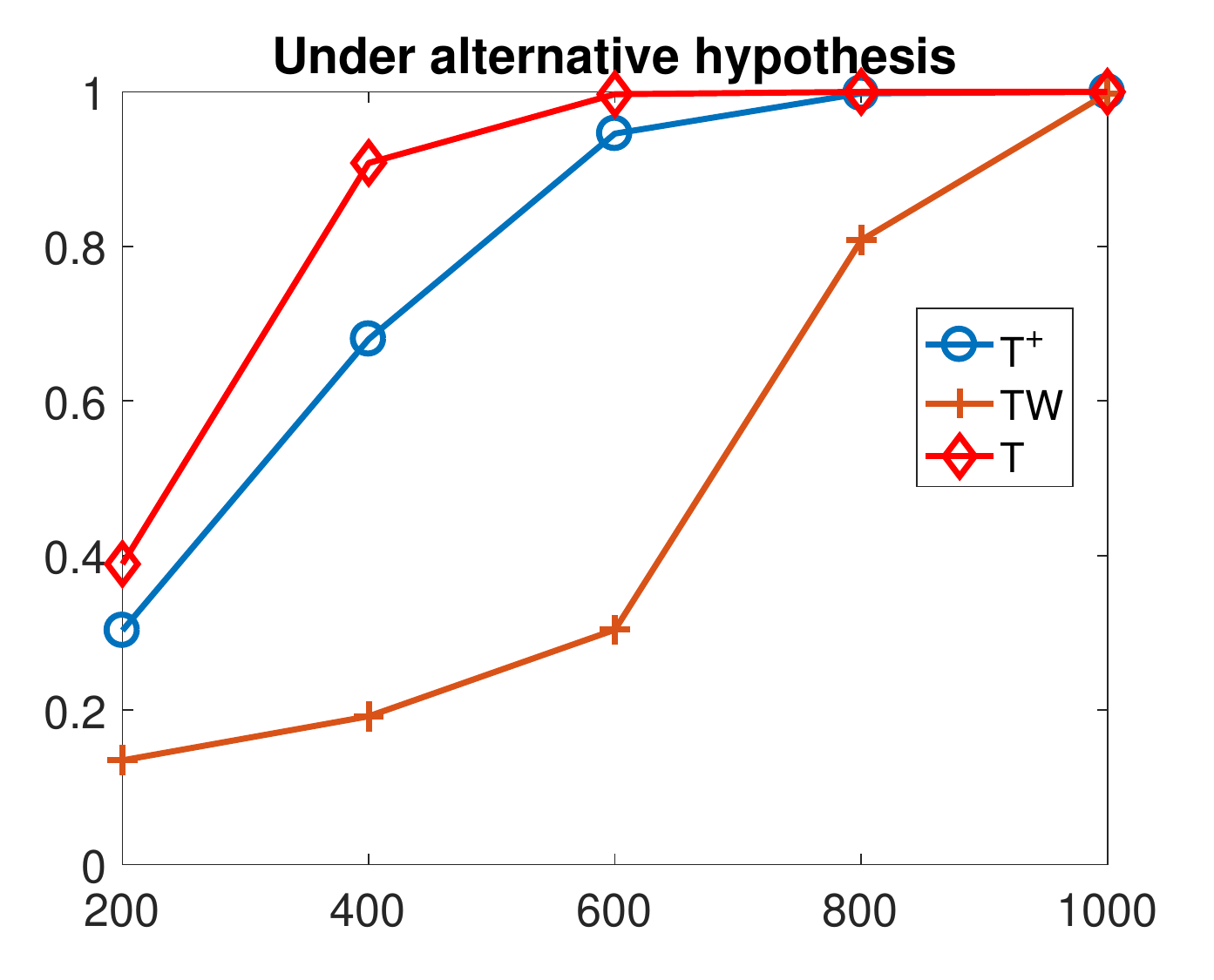}
			
		\end{minipage}
		\caption{ Comparation for the tests with an increasing number of vertices $n$ and $r=1$. }\label{fig2}
	\end{figure}

\begin{figure}[h]
		\centering
		\begin{minipage}[t]{0.48\textwidth}
			\centering
			\includegraphics[height= 6cm,width=6cm]{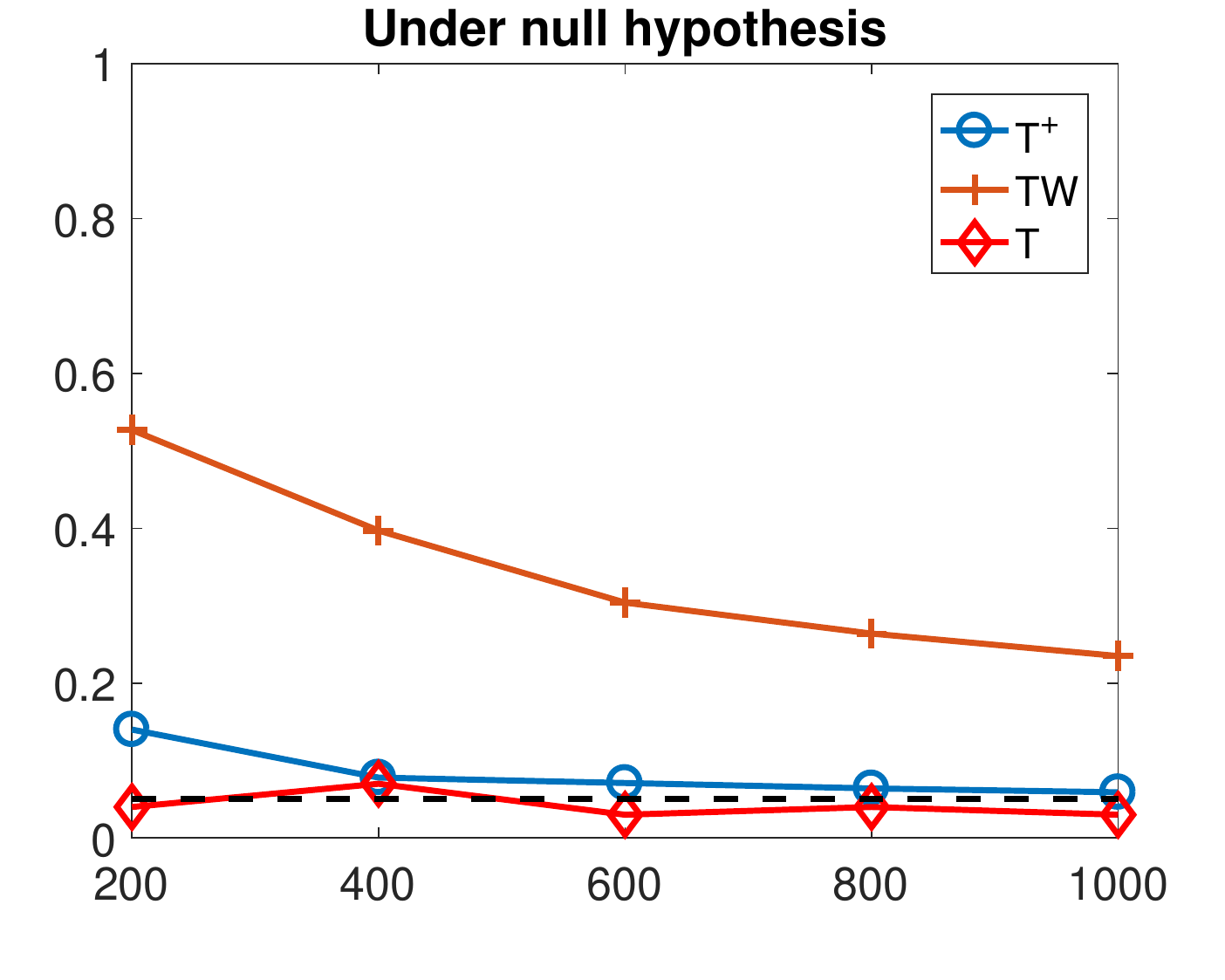}
			
		\end{minipage}
		\begin{minipage}[t]{0.48\textwidth}
			\centering
			\includegraphics[height= 6cm,width=6cm]{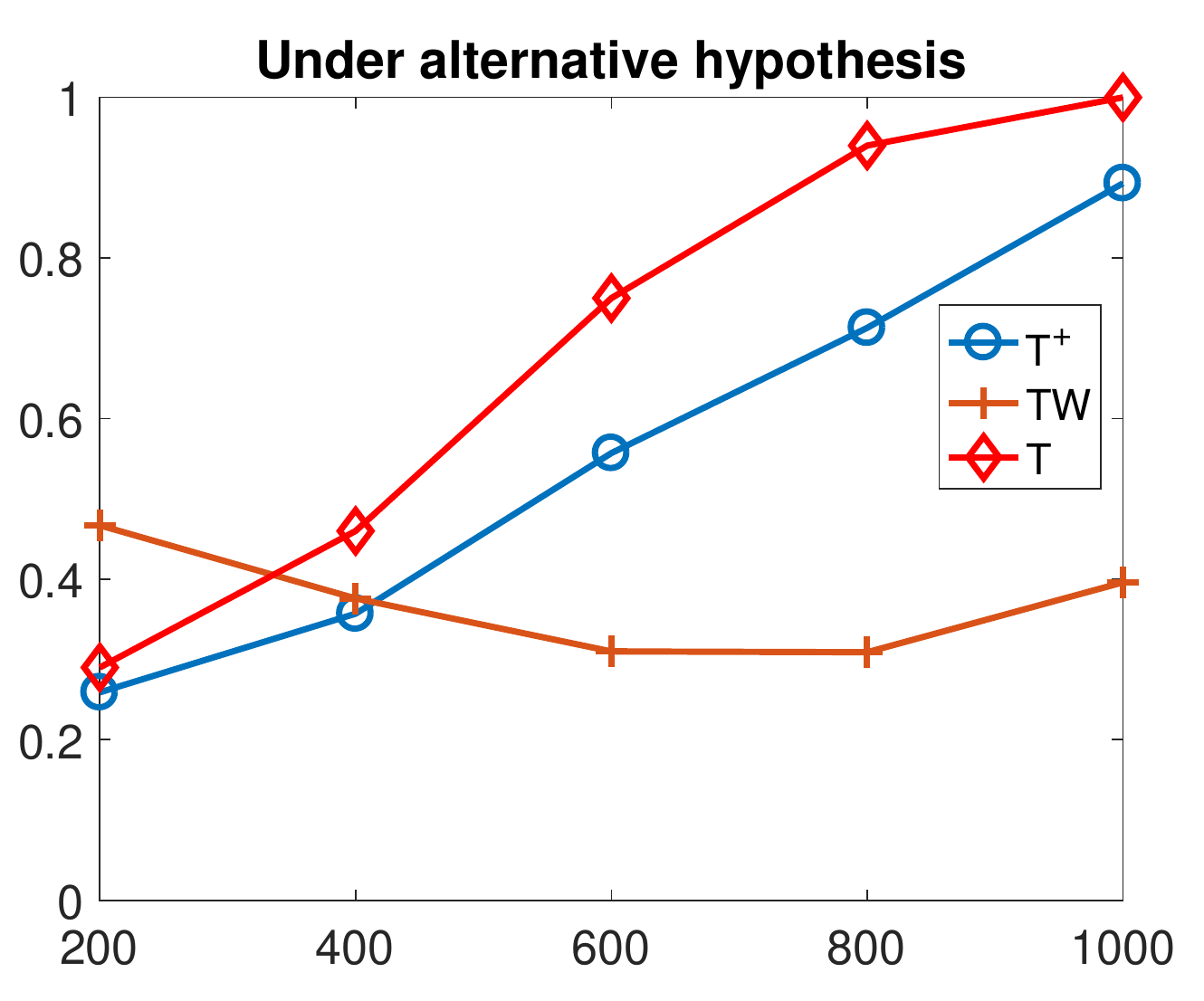}
			
		\end{minipage}
		\caption{ Comparation for the tests with an increasing number of vertices $n$ and $r=1$. }\label{fig3}
	\end{figure}

	We take the same parameter settings as in \cite{WuTwoSampleSBM2022}. The edge probabilities between communities $k$ and $l$ are $B_{1,kl}=0.05r+0.05r\times I\{k = l\}$, $B_{2,kl}=B_{1,kl}+\epsilon r\times I\{k = l\}$, where $r$ controls the sparsity of the network. Under the null hypothesis $\epsilon = 0$, we have $B_1 = B_2$ apparently. Under the alternative hypothesis,  we set  $\epsilon = 0.04$,  which is the same as that in \cite{WuTwoSampleSBM2022}. The membership vector $g$ is generated by sampling each entry independently from $\{1, 2\}$ with equal probability.
	First, we fixed $r=1$ and let $n$ increase from 200 to 1000. Under 1,000 independent replications, the proportion of rejection
	at the nominal significance level of 0.05 can be seen in Figure \ref{fig2}. The dotted
	line for the null hypothesis case corresponds to the significance level of 5\%. It reveals that our test statistic $T$  consistently outperforms $T^{+}$ and $TW$. 
	Next we consider $r=0.5$, all the other parameters are the same as in the setting $r=1$.  It can be see in \ref{fig3} that our test statistic $T$ are more robust to sparse networks than $T^{+}$ and $TW$.   
   \subsection{Type I and Type II errors under the sparse networks}
\begin{figure}[h]
    	\centering
    	\begin{minipage}[t]{0.48\textwidth}
    		\centering
    		\includegraphics[height= 6cm,width=5.8cm]{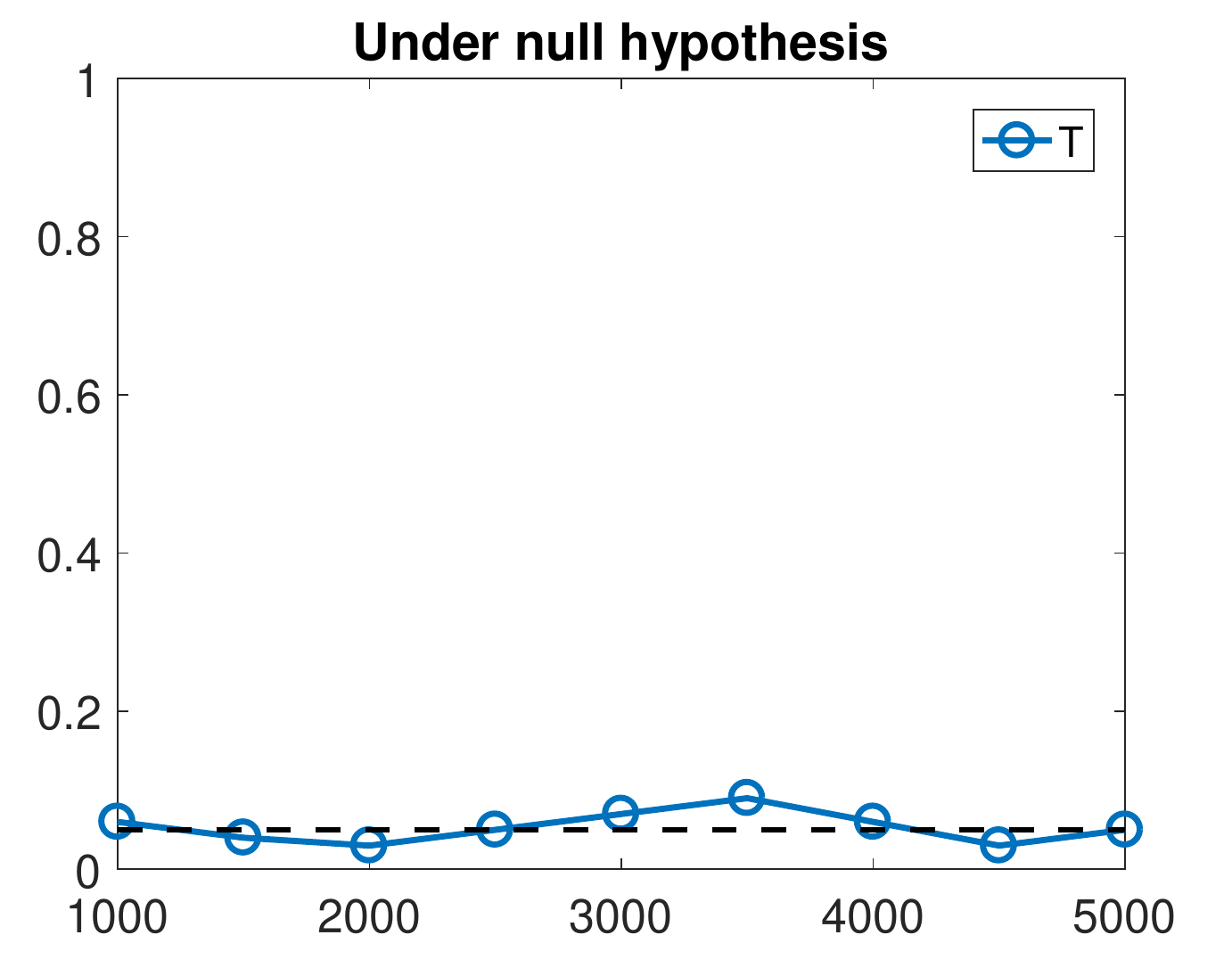}
    		
    	\end{minipage}
    	\begin{minipage}[t]{0.48\textwidth}
    		\centering
    		\includegraphics[height= 6cm,width=5.8cm]{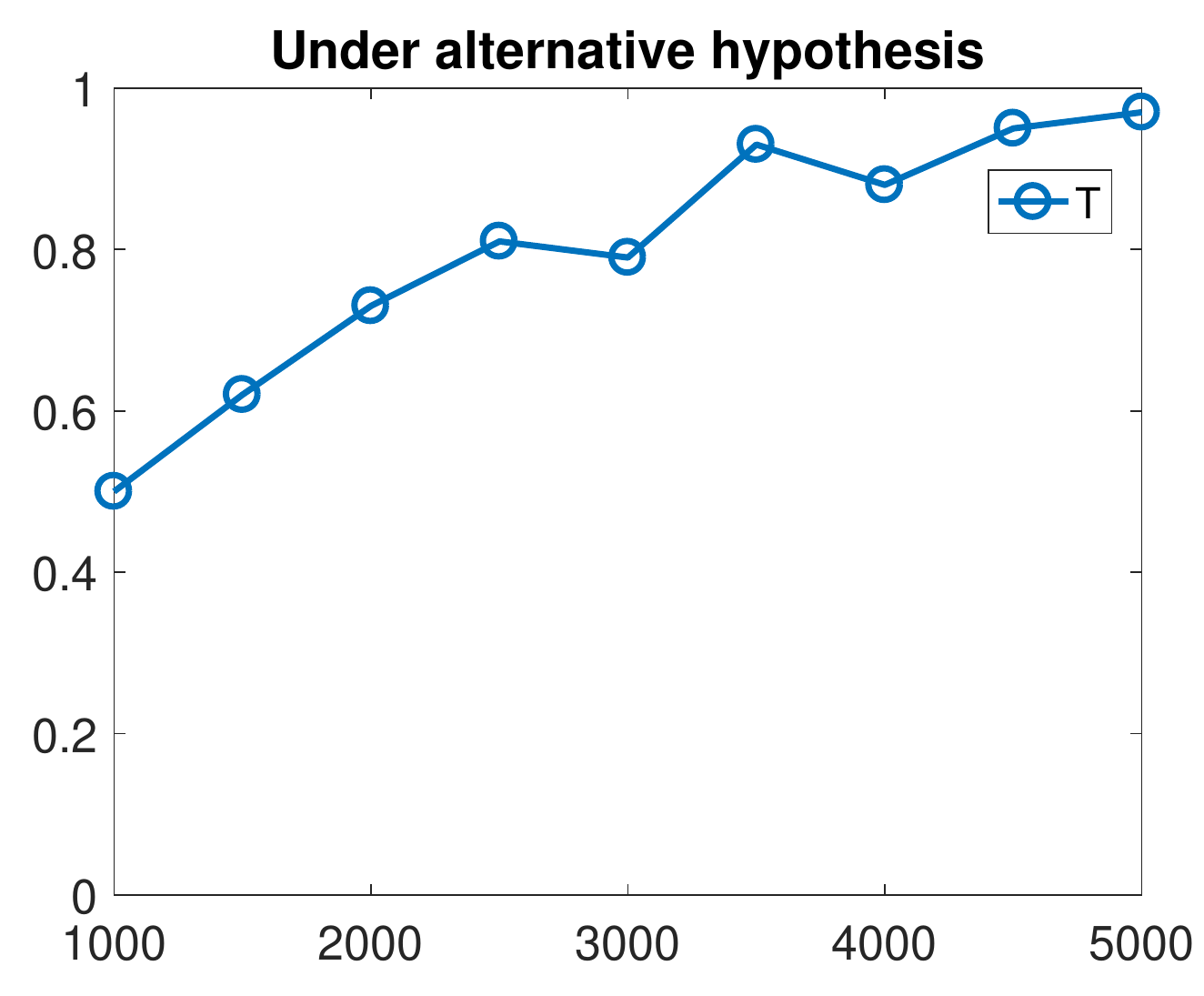}
    		
    	\end{minipage}
    	\caption{ Comparation for the tests with an increasing number of vertices $n$ and $r=1$. }\label{sparse}
    \end{figure}

    In this simulation, we illustrate the performance of our test statistic $T$ since the other two test statistic $T^{+}$ and $TW$ fail in the sparse network settings.
    In the sparse network settings, the edge probabilities between communities $k$ and $l$ are $B_{1,kl}=0.5r+0.5r\times I\{k = l\}$, $B_{2,kl}=B_{1,kl}+\epsilon r\times I\{k = l\}$,  where $r = \frac{\log n}{n}$. Under the null hypothesis $\epsilon = 0$, we have $B_1 = B_2$. Under the alternative hypothesis, we set $\epsilon = 0.4$. The membership vector $g$ is generated by sampling each entry independently from $\{1, 2\}$ with equal probability.
    We let the network size $n$ increases from 1000 to 5000. Under 1,00 independent replications, the proportion of rejection at the nominal significance level of 0.05 can be seen in Figure \ref{sparse}. The dotted
    line for the null hypothesis case corresponds to the significance level of 5\%. It reveals that our test statistic $T$  works well in the sparse network settings.
    
	\section{Real Data Example}\label{rdata}
	In this section we conduct experiments on the real dataset. The  data is collected at the Department of Computer Science at Aarhus University, pertaining to employee interactions. The population of the study included 61 employees who voluntarily participated in the survey, comprising professors, postdoctoral researchers, PhD students and administration staff, out of the total of 142 employees. This dataset including five adjacency matrices which focus on measuring 5 structural variables, namely current working relationships, repeated leisure activities, regularly eating lunch together, co-authorship of a publication, and friendship on Facebook. The relationships between the actors were dichotomous and based on their interactions, without any weighting. The first three variables were assessed through a questionnaire that was distributed online and measured individual assessments made by the actors. The co-authorship relationship was obtained from the online DBLP bibliography database, without any direct interaction with the actors. A custom application was utilized to obtain friendship connections among all individuals who reported having a Facebook account. More details can be found in \cite{rossiEffectiveVisualAnalytics2015}.
	
\begin{figure}[h]
		\centering
		\begin{subfigure}[b]{0.4\textwidth}
			\centering
			\includegraphics[height= 4cm,width=\textwidth]{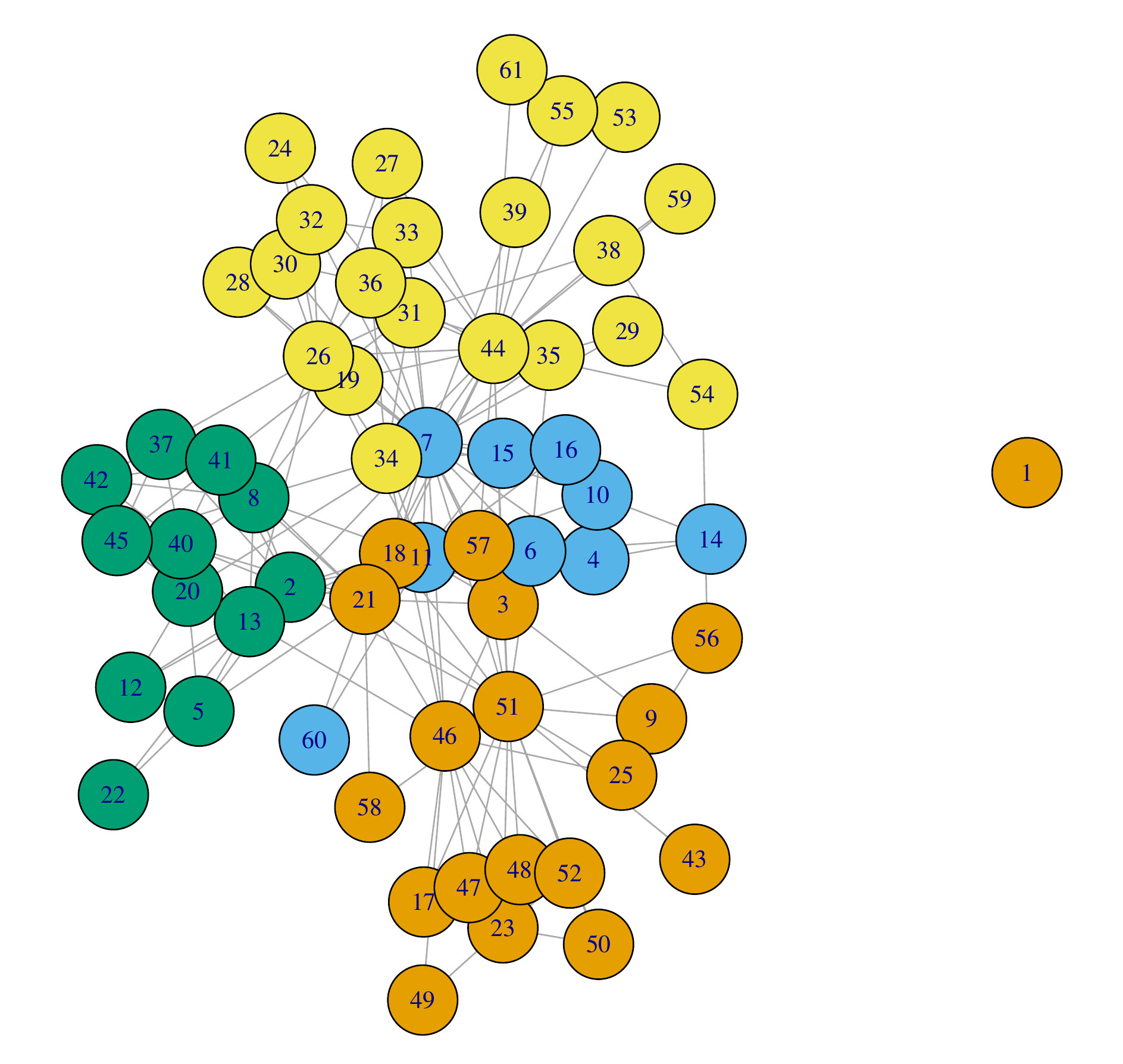}
			\caption{Work}
			\label{}
		\end{subfigure}
		\hfill
		\begin{subfigure}[b]{0.4\textwidth}
			\centering
			\includegraphics[height= 4cm,width=\textwidth]{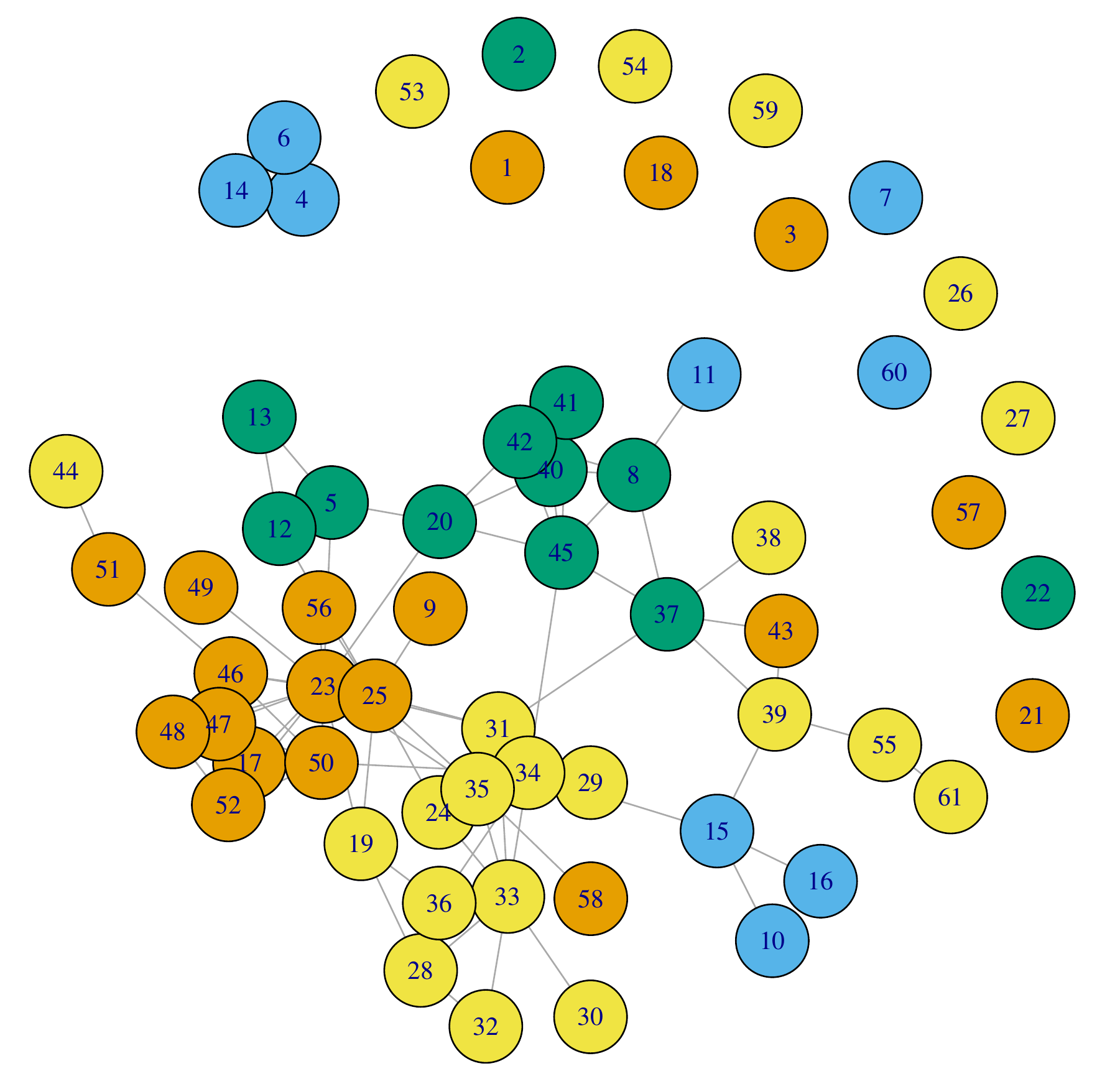}
			\caption{Leisure}
			\label{}
		\end{subfigure}
		\hfill
		\begin{subfigure}[b]{0.4\textwidth}
			\centering
			\includegraphics[height= 4cm, width=\textwidth]{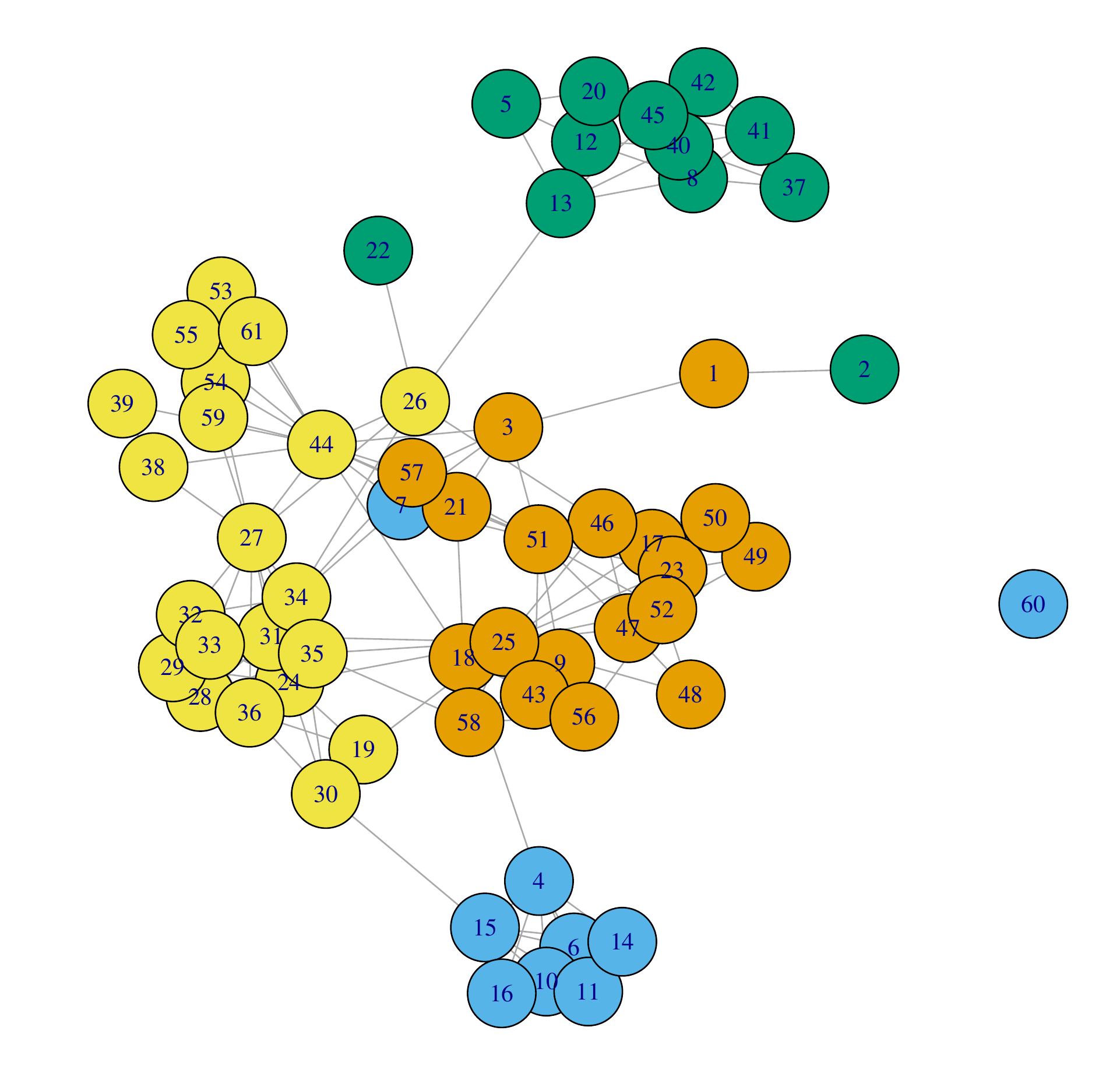}
			\caption{Lunch}
			\label{}
		\end{subfigure}
		\begin{subfigure}[b]{0.4\textwidth}
		\centering
		\includegraphics[height= 4cm, width=\textwidth]{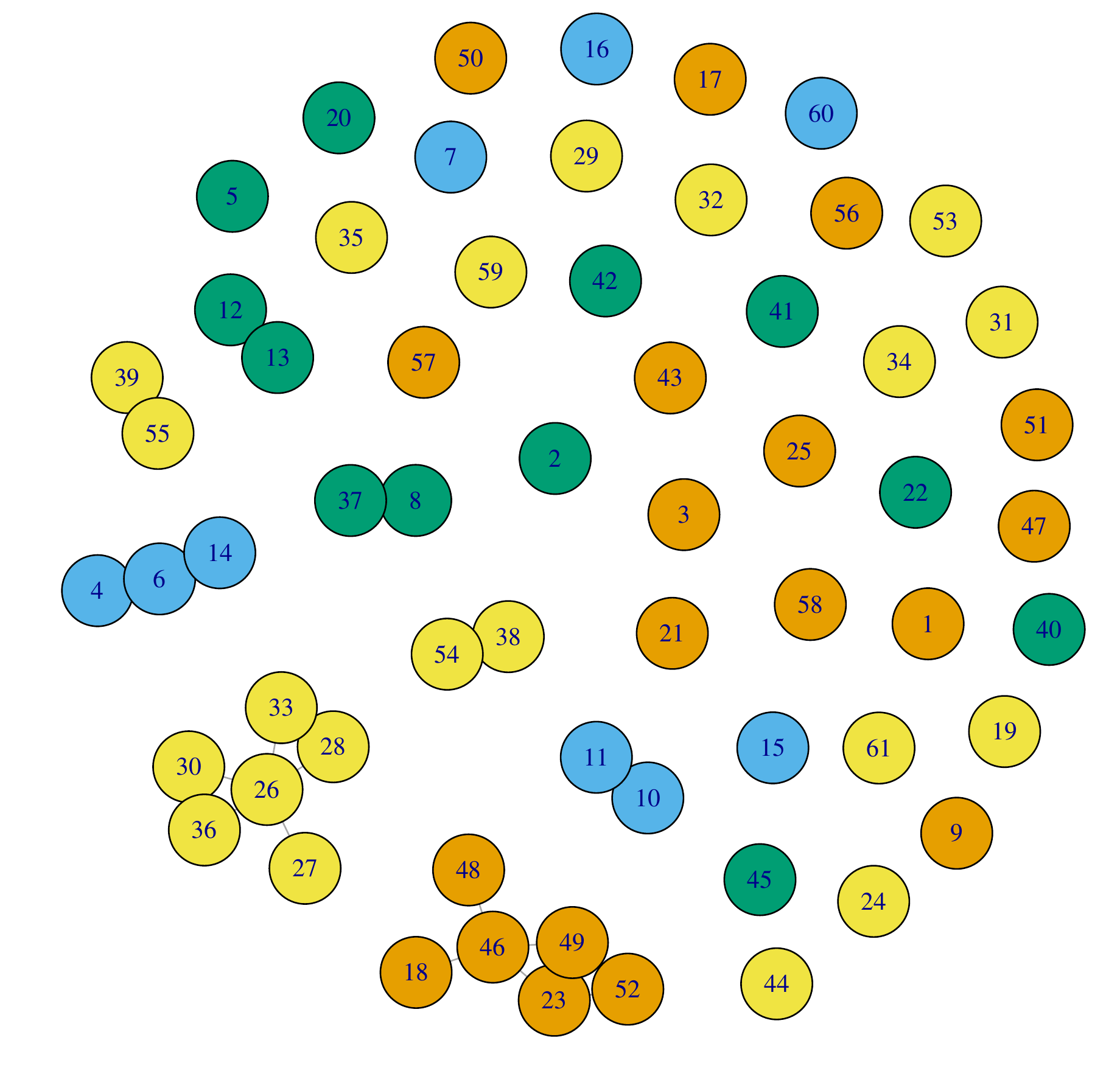}
		\caption{Co-authorship}
		\label{}
	\end{subfigure}
		\begin{subfigure}[b]{0.4\textwidth}
	\centering
	\includegraphics[height= 4cm, width=\textwidth]{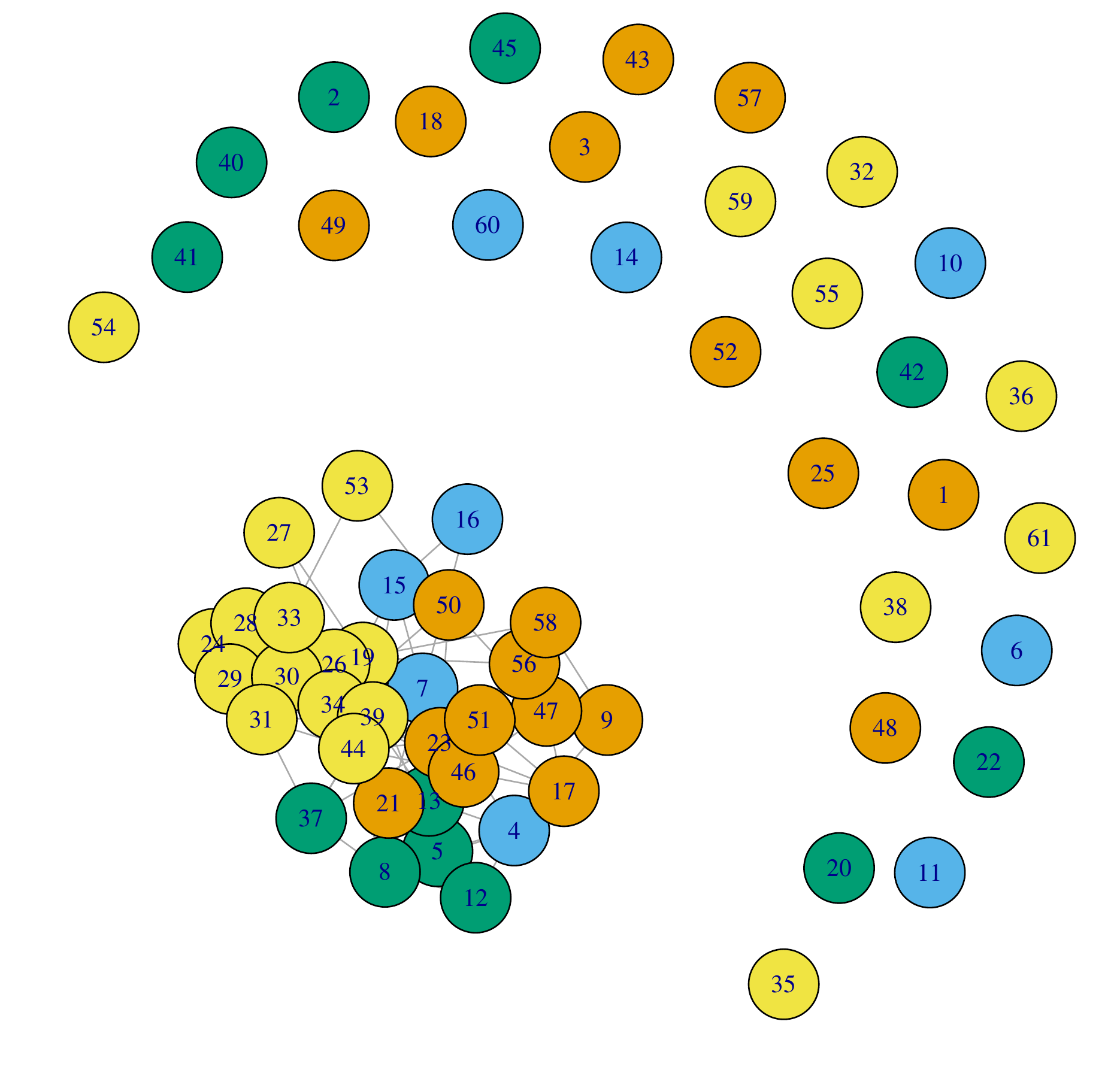}
	\caption{Friendship}
	\label{}
  \end{subfigure}
		\caption{Visualization of the five networks.}
		\label{visul}
	\end{figure}

	We want to characterize the differences between the five community-wise edge probabilities of Work, Leisure, Lunch, Co-authorship, and Friendship, which are denoted as $B_W$, $B_{L_1}$, $B_{L_2}$, $B_{C}$ and $B_F$, respectively. Statistically, we conduct ten two sample testing problems between the five matrices. We use $T$ as the test statistic and calculate the values of  test statistic, the result is listed in Table \ref{tab1}. Since $t_{0.95}$ = 4.79 for the Gumbel distribution, we reject the above ten null hypotheses at the level of 0.05 with strong evidence.     
	For illustration, Figure \ref{visul} shows the five networks, where the users are colored based on the clustering results. 
	
    \begin{table}[h]
    \centering
    \setlength{\tabcolsep}{7mm}
    \caption{\label{tab1} The values of test statistic corresponding to ten two sample testing problems between the five matrices.}
	\begin{tabular}{|l|l|l|l|l|}
		\hline
		Work & 47.05 & 14.99 & 50.19   & 12.95       \\ \hline
		& Leisure & 11.43  & 13.44 & 11.58     \\ \hline
		&         & Lunch & 20.50   & 42.47      \\ \hline
		&         &       & Co-authorship & 22.76      \\ \hline
		&         &       &               & Friendship \\ \hline
	\end{tabular}
    \end{table}

\section{Conclusion}\label{conclusion}
In this paper, we propose a new approach to test whether two sparse networks
are from the same SBM with $K$ communities. We derive the asymptotic null distribution of the test statistic under the null hypothesis and provide the theoretical power under the alternative hypothesis. Real-world networks often exhibit degree heterogeneity, which can be accommodated by extending the proposed method to the degree-corrected SBM. However, when the degree correction parameters are unknown, it becomes challenging to derive the asymptotic distribution because of the complex dependency between the entries of the rescaled adjacency matrix. We leave this for the future work. Simulations and real data examples validate our theoretical results.

\section*{Acknowledgments}
Jiang Hu was supported by National Natural Science Foundation of China (Grant Nos. 12171078 and
11971097).
		

	
\bibliography{MyLibrary}
	
\bibliographystyle{elsarticle-harv}
\end{document}